\documentclass[sigconf]{acmart}

\usepackage{booktabs} 
\usepackage{paralist}
\usepackage[ruled, linesnumbered]{algorithm2e}
\usepackage{amsmath}
\usepackage{url}
\usepackage{multirow}
\usepackage{booktabs}
\usepackage{graphicx}
\usepackage{subfigure} 
\usepackage{bm}
\usepackage{balance}

\def\BibTeX{{\rm B\kern-.05em{\sc i\kern-.025em b}\kern-.08emT\kern-.1667em\lower.7ex\hbox{E}\kern-.125emX}}
    
\begin{document}

\title{ACE-BERT: Adversarial Cross-modal Enhanced BERT for E-commerce Retrieval}


\author{Boxuan Zhang}
\authornote{Corresponding author}
\email{boxuan.zbx@alibaba-in.com}
\affiliation{
  \institution{Alibaba Group}
  \city{Hangzhou}
  \country{China}
}

\author{Chao Wei}
\email{weichao.wc@alibaba-inc.com}
\affiliation{
  \institution{Alibaba Group}
  \city{Hangzhou}
  \country{China}
}

\author{Yan Jin}
\email{yan.jinyan@alibaba-inc.com}
\affiliation{
  \institution{Alibaba Group}
  \city{Hangzhou}
  \country{China}
}

\author{Weiru Zhang}
\email{weiru.zwr@alibaba-inc.com}
\affiliation{
  \institution{Alibaba Group}
  \city{Hangzhou}
  \country{China}
}

%
\renewcommand{\shortauthors}{Zhang et al.}

%
\begin{abstract}
Nowadays on E-commerce platforms, products are presented to the customers with multiple modalities. These multiple modalities are significant for a retrieval system while providing attracted products for customers. Therefore, how to take into account those multiple modalities simultaneously to boost the retrieval performance is crucial. This problem is a huge challenge to us due to the following reasons: (1) the way of extracting patch features with the pre-trained image model (e.g., CNN-based model) has much inductive bias. It is difficult to capture the efficient information from the product image in E-commerce. (2) The heterogeneity of multimodal data makes it challenging to construct the representations of query text and product including title and image in a common subspace. We propose a novel Adversarial Cross-modal Enhanced BERT (ACE-BERT) for efficient E-commerce retrieval. In detail, ACE-BERT leverages the patch features and pixel features as image representation. Thus the Transformer architecture can be applied directly to the raw image sequences. With the pre-trained enhanced BERT as the backbone network, ACE-BERT further adopts adversarial learning by adding a domain classifier to ensure the distribution consistency of different modality representations for the purpose of narrowing down the representation gap between query and product. Experimental results demonstrate that ACE-BERT outperforms the state-of-the-art approaches on the retrieval task. It is remarkable that ACE-BERT has already been deployed in our E-commerce's search engine, leading to $1.46\%$ increase in revenue. 

\end{abstract}

\maketitle

\section{Introduction}
\begin{figure}
\subfigure[]{
    \includegraphics[width=1.55in]{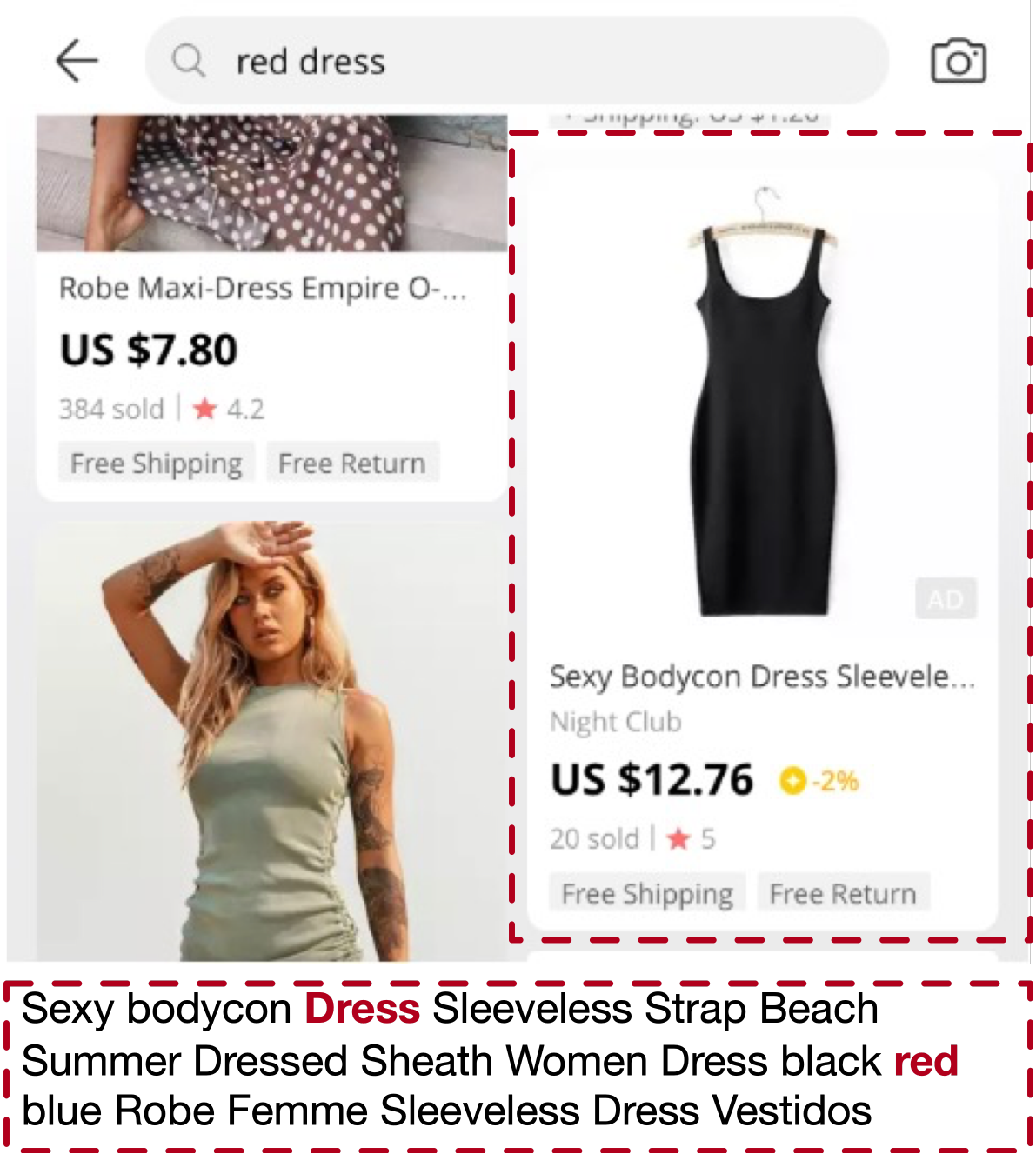}
    \label{title_mismatch}
}
\subfigure[]{
    \includegraphics[width=1.55in]{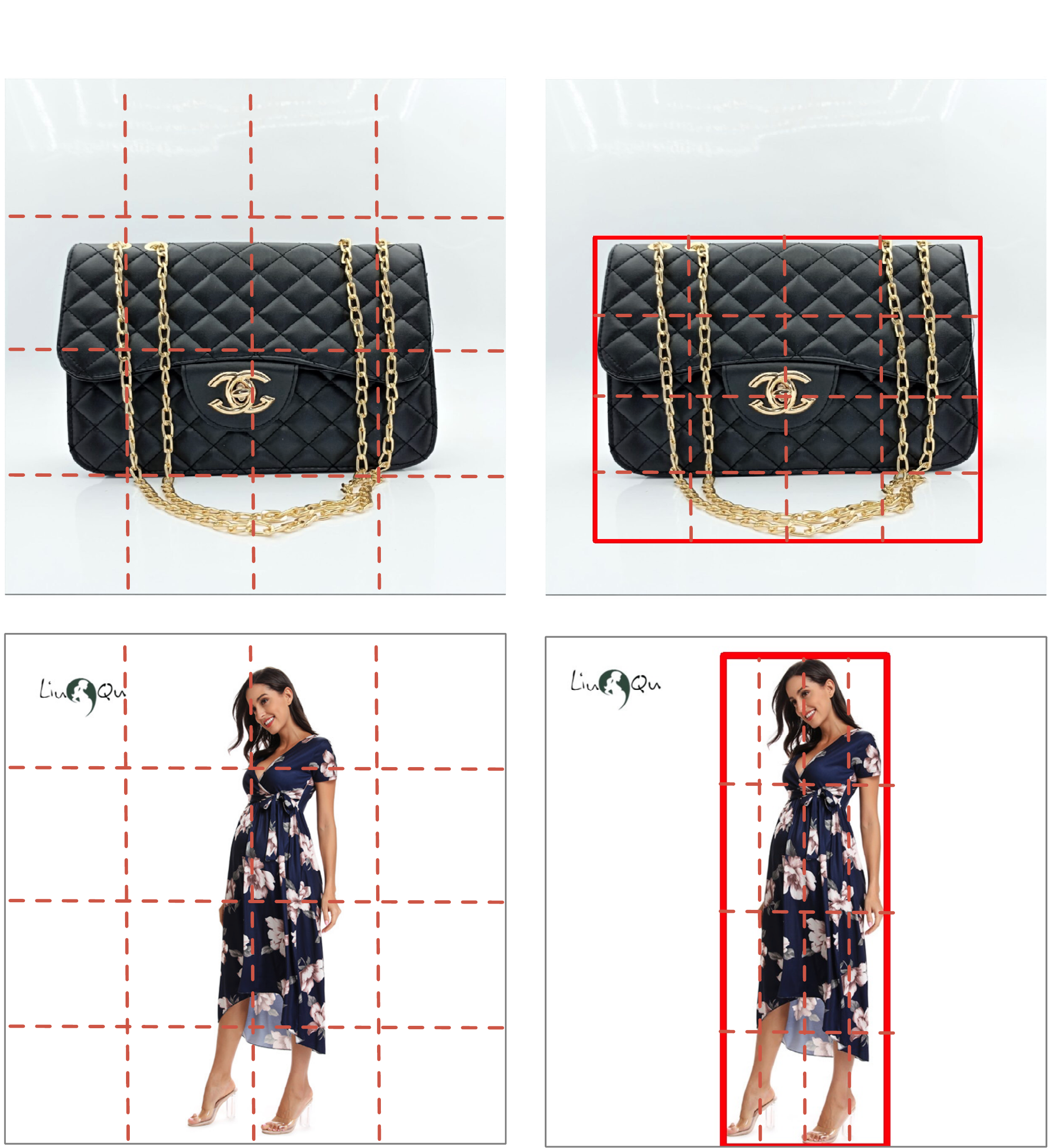}
    \label{image_patch}
}
\caption{(a) the products returned to a user who is looking for "red dress", a mismatched product with its title and image in the dotted bounding boxes is presented to the user, (b) the left part cuts images into patches with the dotted bounding boxes, the right part detects RoIs with red bounding boxes before cutting into patches. } 
\label{product-demo} 
\end{figure}

With the rapid development of Internet, lots of people have got used to purchase on E-Commerce platforms on a global scale, such as Amazon\footnote{www.amazon.com}, Aliexpress\footnote{www.aliexpress.com} and Ebay\footnote{www.ebay.com}. Usually, a user (customer) issues a query text and the relevant products which are consist of multiple modalities (e.g. text, image and video) are presented to the user after being retrieved and ranked by the search engine. As the bottom of the search engine, the performance of the retrieval system is the bottleneck of the whole search engine. 
However, the majority of traditional research on retrieval system is still based on term matching\cite{baeza1999modern}. In the context of E-commerce retrieval, it is significant to understand not only the deep semantic information of the query and product title, but also the multimodal information of the product since: (1) term matching performs well only for the cases that terms of query and product title are matched. (2) The mismatch between query text and product image may seriously impact the user's experience. A example is shown in Fig.\ref{title_mismatch}. When a user is looking for "red dress", a product (in the dotted bounding boxes) with title containing "red dress" is presented to the user. However, there is an obvious mismatch between query and this product image for the reason that current E-commerce retrieval system does not take into account the multimodal information.

Recently, many deep learning algorithms are proposed and successfully applied to the retrieval system. Huang et al.\cite{huang2020embedding} proposed EBR for Facebook search. It is a popular way that applying semantic embeddings to represent query and product by the deep neural network, and then converting the retrieval problem into a nearest neighbor search\cite{johnson2019billion} problem. Therefore, in research field of the retrieval system, most of the previous research was focused on representation learning\cite{bengio2013representation}. BERT\cite{devlin2018bert}, a pre-trained model, learns the deep bidirectional representations from unlabeled text and performs better than other state-of-the-art algorithms. 
Gao et al. proposed FashionBERT\cite{gao2020fashionbert}, which is a extended BERT to address the cross-modal retrieval problem in fashion industry. FashionBERT contributes to retaining the fine-grained information by replacing the RoIs with patches as image tokens and regards the outputs of the pre-trained image model as patch features. This method does not work well in E-commerce context since: (1) the product image usually contains only one object. A sample is shown in Fig.\ref{image_patch}. When the origin image is cut into patches, half of patches shown in left part of Fig.\ref{image_patch} are insignificant. (2) The pre-trained image model based on CNN structure has much more image-specific inductive bias\cite{dosovitskiy2020image} than transformers.


In this paper, we focus on the retrieval system from the perspective of E-commerce. We study how to introduce the multimodal features of the product and then construct the representations of query and product in a common subspace. Different from text and image matching in general cross-modal retrieval, we pay much attention to not only the multimodal fusion of the product, but also the multimodal alignment between query and product embedding. We propose the novel Adversarial Cross-modal Enhanced BERT (ACE-BERT) to solve these problems. Unlike the image features in other multimodal BERT, we propose the RoI-based patch method to extract image tokens. Each product image is detected RoI (i.e., shown in the right part of Fig.\ref{product-demo}) before cut into patches. Besides the patches, we also employ the pixel features to represent image features. In addition, inspired by \cite{wang2017adversarial}, our ACE-BERT employs a domain classifier to play a minimax game with the encoder (i.e., the multimodal BERT). The purpose is to ensure the distribution consistency of the representations generated by the encoder.

To summarize, the contributions are as follows: 
(1) to better extract the information from the product image in an E-commerce-oriented view, we introduce the RoI-based patch method to cut the image into patches. We use the patch-level and pixel-level feature for image representation.
(2) We further employ the adversarial learning to narrow down the representation gap between query and product.
Both large-scale offline experiments and online A/B tests are systematically carried out in our E-commerce's production environment to verify the state-of-the-art performance.


\section{The Pre-trained Enhanced BERT}
In this section, we will detail the pre-training of our ACE-BERT. We refer to this pre-trained cross-modal enhanced BERT as CE-BERT in the following sections. 
 


\subsection{Model Architecture}
\begin{figure*}
\centering
{\includegraphics[width=6in]{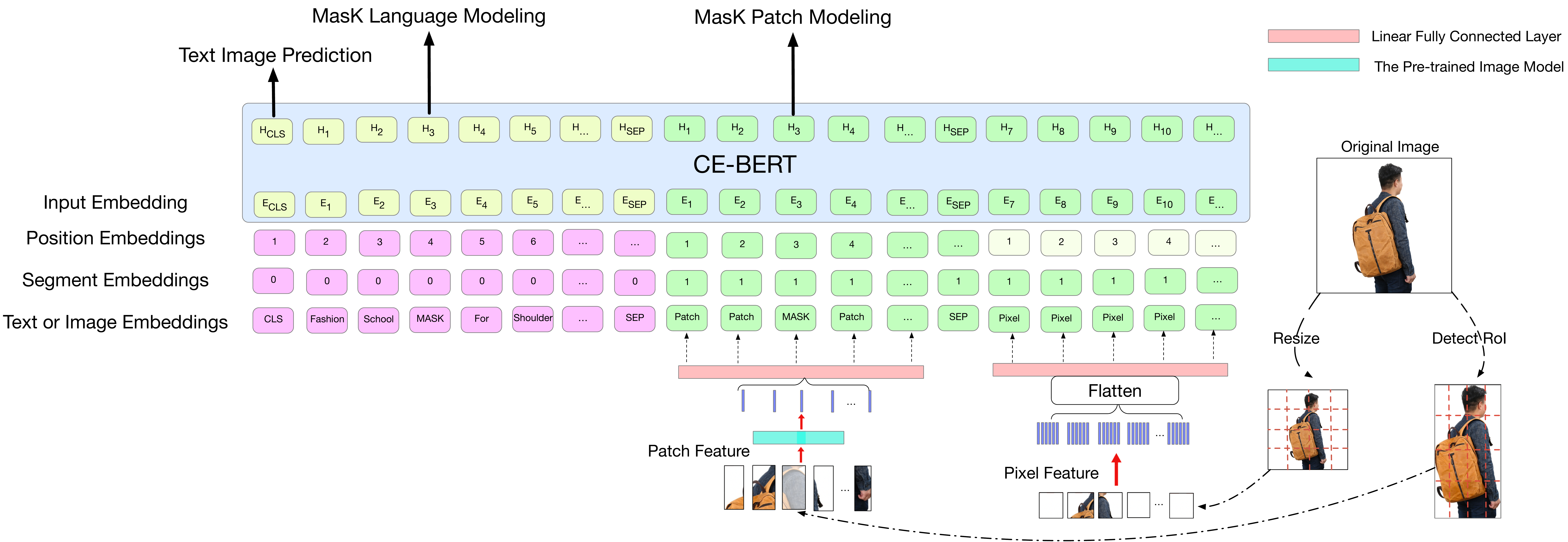}}
\caption{The CE-BERT model architecture}
\label{AM-BERT-model}
\vspace{-4mm}
\end{figure*}

The BERT\cite{devlin2018bert} model was initially proposed by Google. It is consist of multi-layer bidirectional Transformer. Fig.\ref{AM-BERT-model} illustrates the overall architecture of CE-BERT. It extends the BERT by adding image patch features and pixel features which will be reported in the following to the input sequences.

\vspace{2mm} \noindent \textbf{Text Representation:} We follow the standard preprocessing of BERT. The text representation is sum of token embeddings, segment embeddings and position embeddings.

\vspace{2mm} \noindent \textbf{Image Representation:} Different from \cite{gao2020fashionbert},\cite{li2019visualbert}, we have two parts of image representation, patch representation and pixel representation. A special [SEP] token is assigned to separate them.

For patch representation, we firstly detect the object-level RoI from the origin image by Edge Detection Method. As Fig.\ref{image_patch} shows, solid red line denotes the detected RoI, which represents the object-level information of the product. Secondly, we cut the detected RoI into patches. Patch features $x_{p_1}\in\mathbb{R}^{N\times{D_1}} $ are extracted from feature maps of CNN (e.g. Resnet), where $N$ is the number of patches and $D_1$ is the dimensions (2048-d in paper) of patch features. As shown in Fig.\ref{AM-BERT-model}, a linear fully-connected (FC) layer is used to map the patch features $x_{p_1}$ to $x_{patch}\in\mathbb{R}^{N\times{D_2}}$, where $D_2$ is the constant latent vector size of the transformer. $D_2$ is equal to the dimensions of the token embeddings. Finally, patch representation is sum of patch embeddings, segment embeddings and position embeddings.

For pixel representation, to obtain the fixed-length representation for facilitating further computation of transformer, we resize the image to a fixed-size resolution $x_{img}\in\mathbb{R}^{H\times{W}\times{C}}$, where C is the channels of image and (H,W) is the resolution. And then, $x_{img}$ is split into a sequence of patches $x_{p_2}\in\mathbb{R}^{N\times{h}\times{w}\times{C}}$ with the same pixel, where $(h,w)$ is the resolution of patch. For each patch, we reshape it into a sequence of flattened 1D vector. Similar to patch embeddings, a linear fully-connected layer is applied on the flattened vectors $v\in\mathbb{R}^{N\times{(h\cdot{w}\cdot{C})}}$ to obtain final pixel embeddings $x_{pixel}\in\mathbb{R}^{N\times{D_2}}$. Finally, the position embeddings are added to the pixel embeddings to retain ordinal information and the segment embeddings are added to indicate the segment it belongs to.

\subsection{Training Tasks}
We pre-train CE-BERT with three tasks, Masked Language Modeling (MLM), Masked Patch Modeling (MPM) and Text and Image Prediction (TIP). 

The MLM and MPM task are vary similar to the mask task in BERT pre-trianing. We randomly mask the text tokens and image patches, which stimulates CE-BERT to learn the multimodal information by providing the whole image pixel information. We hypothesise that the pixel features provides a priori knowledge without any inductive bias and is beneficial to multimodal fusion. In particular, the MLM task follows the standard processing of BERT, the model is trained under a cross-entropy loss. 

\begin{align}\label{eq.loss_mlm.}
\mathcal{L}_{MLM}= -\sum{y^mlog(y^m)'}
\end{align}
where $(y^m)'$/$y^m$ represents the predicted and true label of MLM task.

The MPM task is supervised to minimize the KL-divergence over two distributions between image patch features, which is specified as follows:

\begin{align}\label{eq.loss_mpm.}
\mathcal{L}_{MPM}= \sum_{s_i\in{M}}{p(s_i)log\frac{p(s_i)}{q(s_i|w_{\backslash{i}}, \theta)}}
\end{align}
where $p(.)$ denotes the distribution of raw patch features, $q(.)$ denotes the output distribution, $M$ denotes the MASK set, $\theta$ denotes the parameters of CE-BERT.

In addition, the TIP task is to predict whether the image and text are from one same product and matched. We model it as a binary classification problem. 

\begin{align}\label{eq.loss_tip.}
\mathcal{L}_{TIP}= - \sum [y \log y' + (1-y)log(1-y')]
\end{align}
where $y'$ represents the predicted label for TIP.

In summary, CE-BERT jointly optimizes three task (i.e. MLM, MPM, TIP) and update for these tasks based on the same mini-batch of instances.

\section{The Adversarial Cross-modal BERT}
In this section, we detail the fine-tuning of ACE-BERT which is consist of two CE-BERTs. At first, we briefly describe how to fine-tune a new retrieval model (called CE-BERT-Finetune) based on CE-BERT. And then, we present ACE-BERT equipped with both additional hot query and adversarial learning.

\subsection{Problem formulation \& CE-BERT-Finetune}

\begin{figure*}
\centering
{\includegraphics[width=6in]{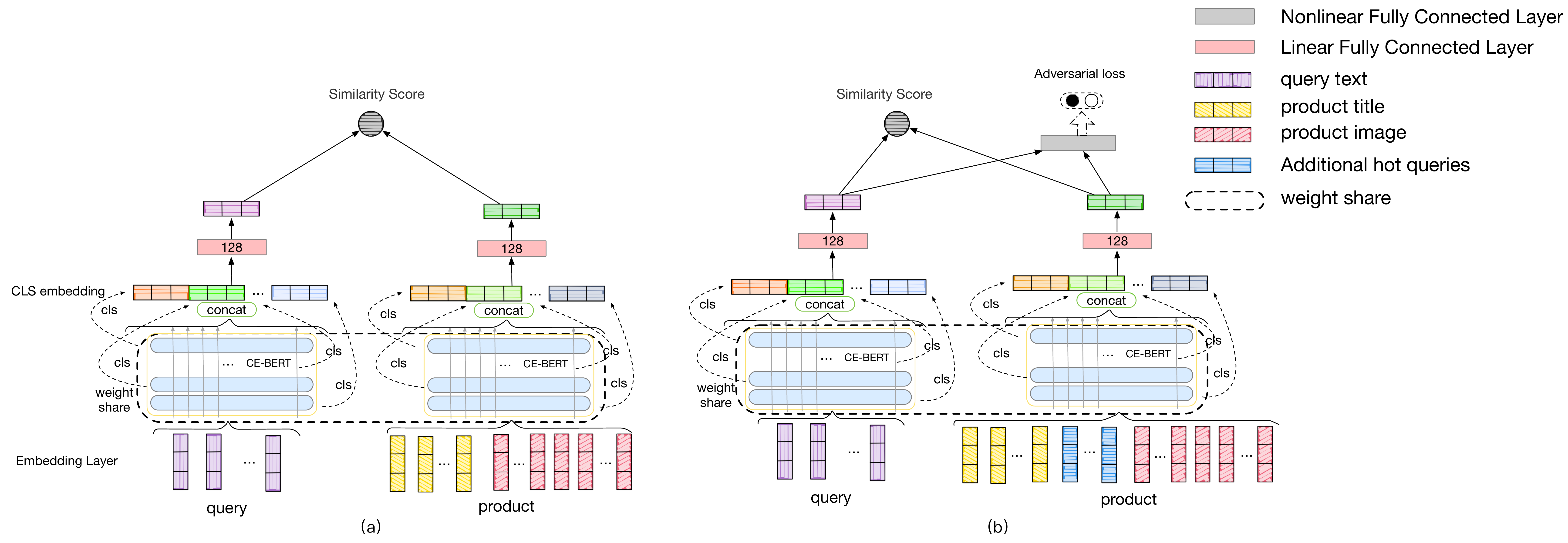}}
\caption{The two fine-tuned cross-modal retrieval models: (a) CE-BERT-Finetune and (b) ACE-BERT }
\label{AC-BERT-model}
\vspace{-4mm}
\end{figure*}

Given a set of queries $\mathcal{Q}=\{q_1,q_2,...,q_s\}$ and their clicked products $\mathcal{A}=\{a_1,a_2,...,a_k\}$, we aim to maximize the conditional likelihood of the clicked product $a_i$ given a query $q_i$. Fig.\ref{AC-BERT-model}(a) shows the main architecture of CE-BERT-Finetune. We firstly encode $q_i$ and $a_i$ as embeddings $v_{q_i}\in\mathbb{R}^d$ and $v_{a_i}\in\mathbb{R}^d$, separately. 

And then, a distance function (e.g. cosine similarity) is applied to $v_{q_i}$ and $v_{a_i}$, 
Similar to \cite{huang2013learning}, we compute the probability from the similarity score of query and product pair through a softmax function:

\begin{align}\label{eq.posterior_probability.}
P(v_{a_i}|v_{q_i})=\frac{exp(\gamma Distance(v_{q_i},v_{a_i}))}{\sum_{j=1}^{w} exp(\gamma Distance(v_{q_i}, v_{a_j}))}
\end{align}

where $\gamma$ denotes a smoothing parameter and we train the Semantic Matching Task to minimize the following bidirectional log loss:

\begin{small}
\begin{align}\label{eq.logloss}
\mathcal{L}_{main}=-\frac{1}{W} (\sum_{i=1}^{W} log P(v_{a_i}^+|v_{q_i}) + \sum_{i=1}^{W} log P(v_{q_i}^+|v_{a_i}))
\end{align}

where $W$ is the number of training instances and $P(v_{q_i}^+|v_{a_i})$ is symmetrical to $P(v_{a_i}^+|v_{q_i})$.
\end{small}

\subsection{Incorporating Features into the Model}
Considering the large-scale and low-latency of the retrieval system in E-Commerce, we need to decouple representation of query and products independently. Thus, product embeddings can be inferred and cached in engine memory in advance. But this change leads to a big reduction in the performance of the retrieval model based on the BERT. The reason should be that compared with BERT in pre-training, the decoupled input sequence of the query and product lacks interaction during fine-tuning. 

\noindent \textbf{Additional Hot Query:} as we can see in Fig.\ref{AC-BERT-model}(b), besides the product title and image,  we employ the additional hot query separated by [SEP] to represent product. The motivation is to let the three features interact with each other in the multi-layer transformers of the product side continuously in training. Thanks to the parameter sharing between query side and product side. It benefits both the representations of query and product. In addition, it is noteworthy that product embeddings can be pre-computed offline and makes no difference to the latency of online serving. 

Here we describe additional hot query in details. When a user submits a query, we collect products that the user clicked, and then count the number of clicks for each query. Finally we use the (at most) top 10 frequent queries as the additional hot query.

\subsection{Adversarial Learning\&Training}
In this section, we will detail how to use adversarial learning to boost the performance of the retrieval task and describe the training procedure of ACE-BERT.

\subsubsection{Why using Adversarial Learning?}
In this paper, we focus on the matching problem between query text and product. we need to represent query with only text sequence while representing product with both text and image sequence in such a scene. That is why the representation gap between query and product is severe. And the condition is getting worse for the reason that we bring the additional hot query in product side leading to a deep representation gap. Therefore, we propose ACE-BERT which takes advantage of adversarial learning to bridge the gap between different modalities. And then, ACE-BERT projects the input features from different modalities into a common subspace.

\subsubsection{Objective \& Training Detail} 
Inspired by \cite{wang2017adversarial}, we define a domain classifier $D$ to detect the domain sources (i.e. query and product) when given an unknown embedding generated by encoder. So the objective is a cross-entropy loss:

\begin{small}
\begin{equation}\label{eq.domain_classifier_loss}
    \mathcal{L}_{adv} = -\frac{1}{W} \sum_{i=1}^{W} [log D(v_{q_i};\theta_D) + log(1 - D(v_{a_i}; \theta_D))]
\end{equation}
\end{small}

where $\bm{x_i}$ is a one-hot vector and denotes the ground-truth domain label of each instance, $\theta_D$ denotes the parameters of $D$.

As shown in Fig.\ref{AC-BERT-model}(b), we refer to the domain classifier as discriminator, which is similar to the discriminator in GAN\cite{goodfellow2014generative}. A minimax two-player game is conducted between the discriminator and encoder. Formally, we incorporate $\mathcal{L}_{main}$ and $\mathcal{L}_{adv}$ to jointly train the model:

\begin{align}\label{eq.adv_optimize}
    (\hat{\theta_{q}}, \hat{\theta_{a}}) = \mathop{\arg\min}_{\theta_q,\theta_a} (\mathcal{L}_{main} - \mathcal{L}_{adv}) \\
    (\hat{\theta_{D}}) = \mathop{\arg\max}_{\theta_D} (\mathcal{L}_{main} - \mathcal{L}_{adv})
\end{align}

The ACE-BERT training procedure is summarized as follow. For the first $k$ iterations, we update the parameters $\theta_q$ and $\theta_a$ by descending the stochastic gradients with the parameters $\theta_D$ of discriminator are fixed. Then in turn, the parameters $\theta_D$ is updated by ascending the gradients with the parameters $\theta_q$ and $\theta_a$ are fixed. The above alternate training procedure is repeated until convergence. 


\section{Experiments}
In this section, we conduct both offline and online experiments to prove the outperformance of ACE-BERT. We provide details of our implementation and experimental setup to help reproduce the findings in this work. We cannot release datasets due to the business secret and privacy issues. 

\subsection{Datasets}
The datasets are collected from our E-commerce platform.
For pre-training, we collect about 4.7 million high-quality products with title and main image as positives (called PT dataset). The ratio of positives to negatives that randomly sampled is kept to 1:3, which is determined by cross validation in preliminary offline experiments. For fine-tuning, we collect over 25 million users' click-through information from search logs (called FT dataset). In particular, we only sample the users' clicked results as positives. And negatives are sampled online in each mini-batch updated during the model training. As mentioned in \cite{huang2020embedding}, blending easy and hard negatives in training is advantageous. We propose the mixed negative sampling in mini-batch which will be reported in the following section. For offline evaluation on the retrieval task, we sample the test set from the users' click-through records in next days.

\subsection{Compared Algorithms}
The compared algorithms in this experiments are as follows:
\begin{itemize}
\item \textit{\textbf{TwinBERT\cite{lu2020twinbert}}}: In this experiment, we use the verison of TwinBERT$_{cos}$ for the reason that TwinBERT$_{res}$ cannot work easily with our ANN algorithm. 

\item \textit{\textbf{FashionBERT-Finetune\cite{gao2020fashionbert}}}: The pre-trained FashionBERT is not evaluated with cross-modal retrieval, we thus fine-tune a new retrieval model with the pre-trained FashionBERT.

\item \textit{\textbf{CE-BERT-Finetune}}: As mentioned before, we fine-tune a new retrieval model based on the pre-trained CE-BERT.

\item \textit{\textbf{ACE-BERT}}: Different from CE-BERT-Finetune, ACE-BERT proposed by us is equipped with the addition hot query and the adversarial learning.
\end{itemize}

We use the Group Area Under the receiver operating characteristic Curve (GAUC)\cite{bradley1997use} and Recall@K (k=10, 50, 100) as evaluation metric in offline experiments. 

We randomly sample a query set. For each query, we construct a target set $T=\{t_1,t_2,...,t_M\}$. And a evaluation set $E=\{e1,e2,...,e_N\}$ is returned by the model. Then the Recall@K is defined as: 

\begin{align*}
    Recall@K = \frac{\sum_{i=1}^n f(e_i)}{M} &&
    f(x)=\begin{cases}
            1 & x_i\in T \\
            0 & else
          \end{cases}
\end{align*}

where $M$ is the size of $T$. $T$ consists of the most relevant products to given query based on certain criteria. In this experiment, the target set is constructed by considering user click log. In particular for calculating the Recall@K, the embeddings of queries and products are generated by models. And then for each query, the ANN search is used to find the top-k products.

\begin{table*}[h]
    \centering
    \caption{Offline comparison results}
    \begin{tabular}{cccccc}
        \hline
        ID&Description & GAUC (Gain) & Recall@10 & Recall@50 & Recall@100 \\
        \hline
        1&TwinBERT & 0.7695 & 3.62\% & 10.16\% & 14.92\% \\
        2&FashionBERT-Finetune & 0.7768 & 3.70\% & 10.33\% & 15.16\% \\
        3&CE-BERT-Finetune & 0.7917 & 3.90\% & 10.89\% & 15.90\%  \\
        4&ACE-BERT & 0.7966 & 5.21\% & 13.61\% & 19.29\% \\
        \hline
    \end{tabular}
    \label{table.methods_offline_res}
\end{table*}

\begin{table}[h]
    \centering
    \caption{Results for online comparisons}
    \begin{tabular}{cccc}
        \hline
         Control & Treatment & CTR & Revenue \\
        \hline
        TwinBERT & FashionBERT-Finetune  & 0.07\% & 0.08\% \\
        TwinBERT  & CE-BERT-Finetune  & 0.32\% & 0.49\% \\ 
        TwinBERT  & ACE-BERT & 0.78\% & 1.46\% \\
        \hline
    \end{tabular}
    \label{table.methods_online_res}
\end{table}

\subsection{Effective of Negative Sampling}
\begin{figure}
\centering 
\includegraphics[width=3in]{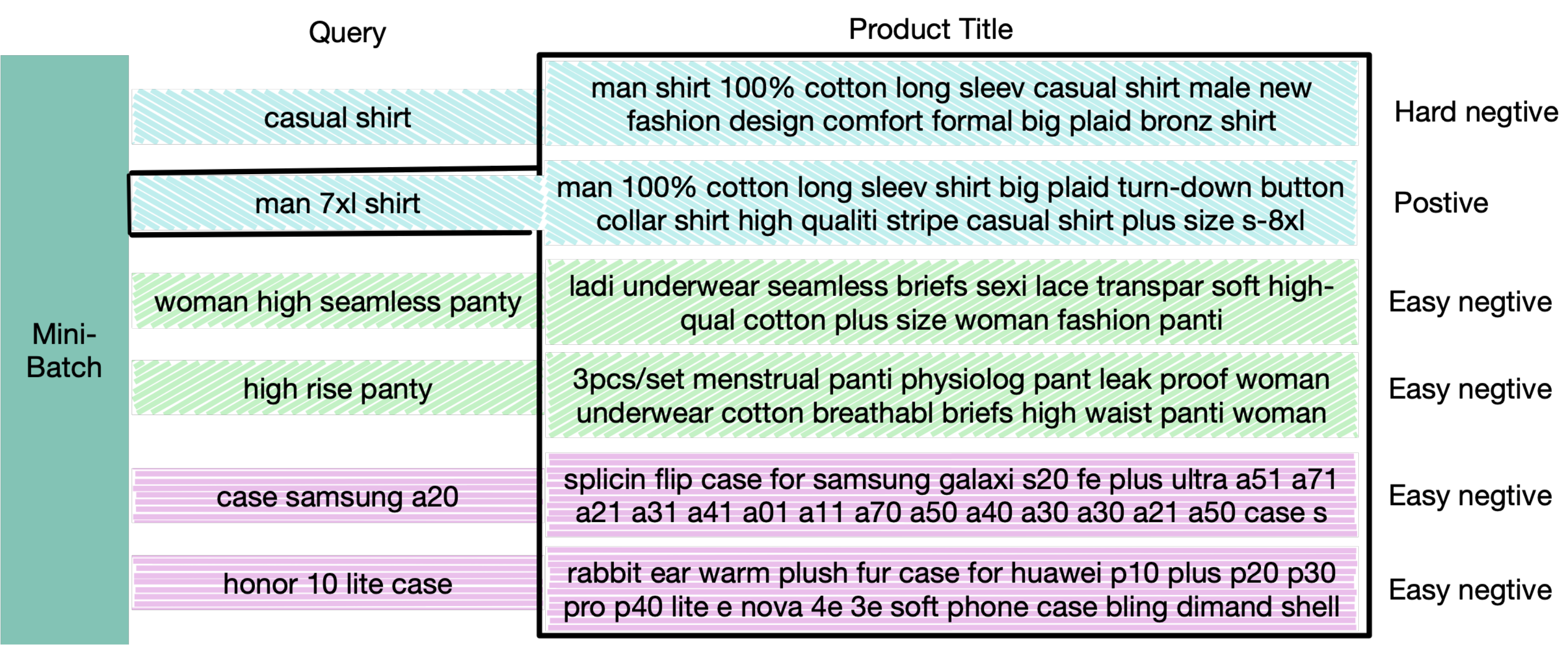}
\caption{The mixed negative sampling in mini-batch, the products' title in black bounding boxes are training instances for "main 7xl shirt".} \label{batch-wise} 
\end{figure}

In this section we will detail the mixed negative sampling in mini-batch. A sample is shown in Fig.\ref{batch-wise}. The mini-batch is consist of positive query and product pairs. For the query of each positive pair in a mini-batch, we use the other product samples as the negatives of the query. This provides a large amount of negatives whose size is coupled with the mini-batch size.

\begin{figure}
\centering 
{\includegraphics[width=2.0 in]{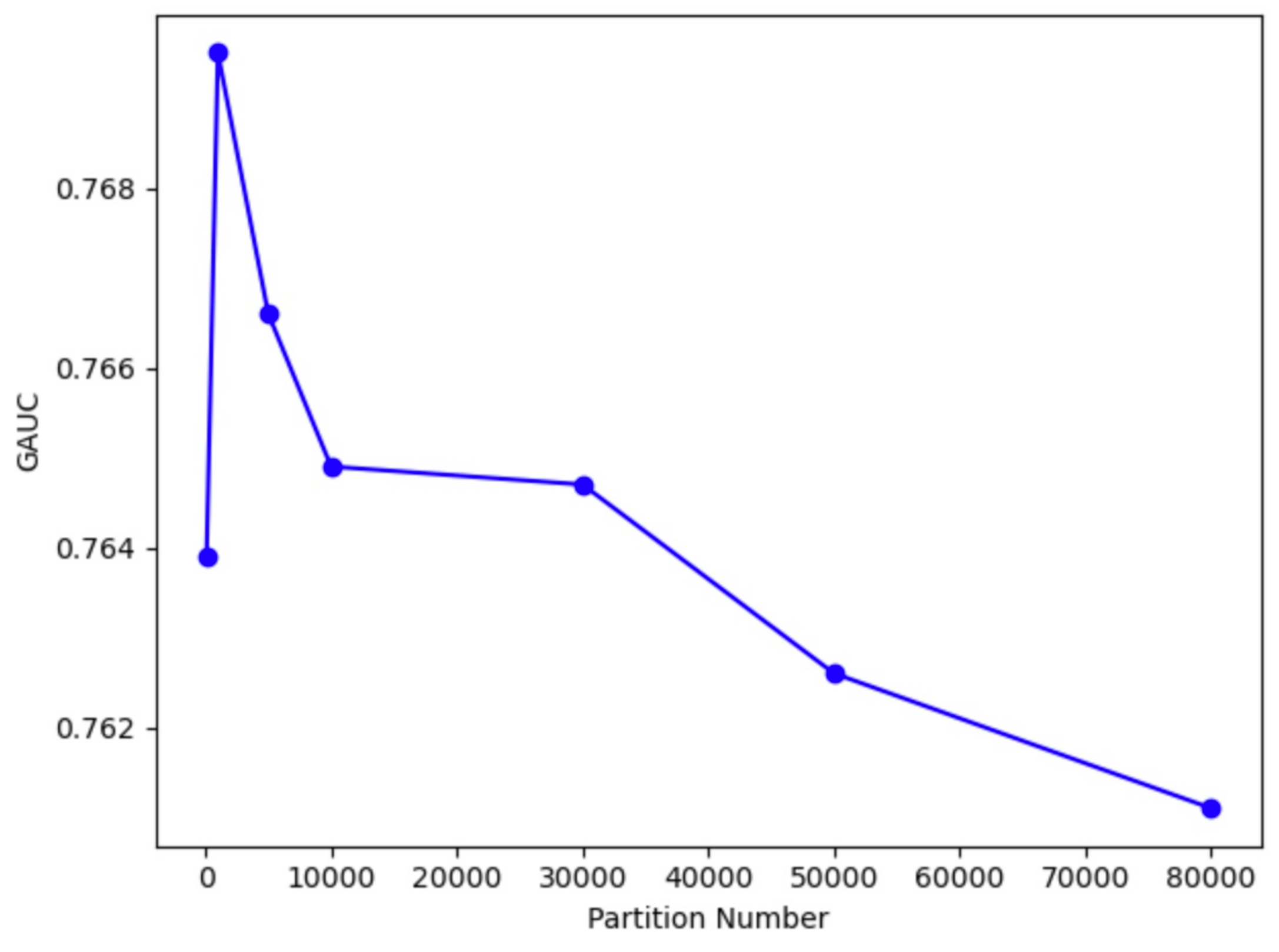}}
\caption{The performance on the test set of the FT dataset when varying the partition number } \label{hard_negative} 
\end{figure}

However, the improvements gains are not significant when the mini-batch size increases continuously. The reason is that these negatives are randomly sampled. When reaching saturation point, those negatives can not provide efficient and effective knowledge for the model learning. From the perspective of the contrastive learning, both quality and quantity of the negatives are important for the model performance. So we randomly split the FT dataset into partitions. For each partition, the positives are clustered according to the product category. We hypothesize that the hard negatives which has the same product category with the positives are beneficial to the model to learn the significant difference among the products. The partition number and the proportion of the hard negatives in mini-batch are inverse ratio. Thus, we adjust the partition number to find its best value. Fig.\ref{hard_negative} shows the results of experiments by setting different partition numbers. According to the results, we set partition number to 1000 for the fine-tuning retrieval task to achieve the best performance. 
In this paper, We try to keep a balance between the limit of our GPU memory size and the model training speed. We thus set the mini-batch size as 384. 

\subsection{Evaluation on Retrieval Task}
\subsubsection{Experiment setup}
Firstly, we initialize the parameters of FashionBERT and CE-BERT with the parameters of the pre-trained BERT and train them on the PT dataset with the settings of hyper-parameters: (1) the 4-layer BERT model is chosen for the limited GPU resource. (2) The maximum text sequence length is set to 512 and the maximum image sequence length is set to 16 and 49 for patch features and pixel features, respectively. For patch features, ResNeXt101\cite{xie2017aggregated} is adopted to extract a 2048d embedding for each patch. For pixel features, the image is resize to $224\times224$ and each patch has $32\times32$ input size. (3) Adam is used as the optimizer with the learning rate set to $2e-5$ and warmed up at the first $10000$ steps. Other hyper-parameters are adopted from BERT pre-training.

Secondly, we evaluate those methods on the retrieval task. Based on above pre-trained models, several new retrieval models are fine-tuned with the following hyper-parameters: (1) the mini-batch size is set to 384 and Adam is used as the optimizer, with learning rate set to $5e-5$ (2) The embedding size of query and product are set to 128. (3) the training will stop when the max number of epochs is reached either early stop to avoid overfitting. Above training is done on five P100 GPUs.

\subsubsection{Comparison Analysis}
We compare these algorithms on the both offline and online. In the offline setting, the GAUC is calculated on the test set in next days. And for Recall@K (k=10, 50, 100), the query set and target set have 100000 randomly sampled queries and 3.4 million clicked products, respectively. On the other hand, the large-scale online A/B tests are conducted to compare those methods. Totally over 100 million search impressions are collected in our A/B tests and search traffic is equally split into the control and treatment groups.

The results are summarized in Table \ref{table.methods_offline_res} and \ref{table.methods_online_res}, respectively. Multimedia data helps to improve E-commerce retrieval performance by comparing FashionBERT-Finetune to TwinBERT. Furthermore, it is found that CE-BERT-Finetune benefits more from patch features and pixel features than from only patch features compared with FashionBERT-Finetune and CE-BERT-Finetune. ACE-BERT achieves the best performance on both GAUC and Recall@K metric. The reason should be that, although the multimodal features help to understand the product in depth, the downstream cross-modal retrieval task will leave a representation gap between query and product. We apply the adversarial learning on ACE-BERT and thus improves its performance. 

From Table \ref{table.methods_online_res}, we can see both CE-BERT-Finetune and ACE-BERT outperform the FashionBERT. ACE-BERT achieves significantly gains over TwinBERT. These results are consistent with the offline results. In addition,  FashionBERT-Finetune shows no significant gains. 

\subsection{Dissection of CE-BERT}
As mentioned above, CE-BERT-Finetune outperforms FashionBERT-Finetune on both GAUC and Recall@K metrics. Here we analyze detailed results to find out whether the pixel features in CE-BERT-Finetune are truly work. 

We compare the pre-trained multimodal BERTs with different image representations. For the sake of fairness, we pre-train a series of multimodal BERTs by using different image representations and fine-tune them on the same FT dataset with the same hyper-parameters. Finally, we evaluate the effectiveness of these models on GAUC and Recall@100 metrics.

\begin{table}[h]
    \centering
    \caption{Evaluation results of the multimodal BERTs with different image representations}
    \begin{tabular}{p{5cm}cc}
        \hline
        Image Representation & GAUC & Recall@100 \\
        \hline
        The Patch Feature  & 0.7768 &  15.16\%\\
        The RoI-based Patch Feature & 0.7816 & 15.30\% \\
        The RoI-based Patch and Pixel Feature & 0.7917 & 15.90\% \\ 
        \hline
    \end{tabular}
    \label{table.pretrain_ablation_res}
\end{table}

The evaluation results are shown in Table \ref{table.pretrain_ablation_res}. Compared with the patch feature used in FashionBERT, the RoI-based patch feature achieves the better performance. In addition, the RoI-based patch and pixel feature used in CE-BERT achieves the best performance, indicating that the way of extracting the image feature used in our model is meaningful for E-commerce retrieval.

\subsection{Ablation Study on ACE-BERT}
Both offline and online experiments confirm that ACE-BERT helps to improve the performance of the retrieval task. In detail, ACE-BERT has the following optimizations: (1) the additional hot query is proposed to enhance the interaction between two encoders of ACE-BERT. (2) Adversarial learning is proposed to align the distribution of representations across domains (i.e., query and product). Thus in this section, we conduct an ablation study of these optimizations. 

\begin{table}[ht]
    \centering
    \caption{The performance of the ablation study for ACE-BERT }
    \begin{tabular}{cccc}
        \hline
        ID&Description & GAUC & Recall@100 \\
        \hline
        1 & No Additional Optimization & 0.7917 & 15.90\% \\ 
        2 & Add Additional Hot Query & 0.7934 & 18.85\% \\
        3 & Add Adversarial Learning & 0.7966 & 19.29\%\\
        \hline
    \end{tabular}
    \label{table.finetune_ablation_study}
\end{table}

In particular, we reuse the parameters of CE-BERT and fine-tune a series of the retrieval models. The offline results are summarized in Table \ref{table.finetune_ablation_study}. 
The model with additional hot query in Setting 2 achieves significant Recall@100 improvements compared to Setting 1, which proves its superiority in the retrieval task. Furthermore, the Setting 3 with adversarial learning achieves significantly gains over Setting 2, which indicates that the fine-tuning step can bridge representation gap across domain by introducing adversarial learning.

\begin{figure}
\centering 
\includegraphics[width=2in]{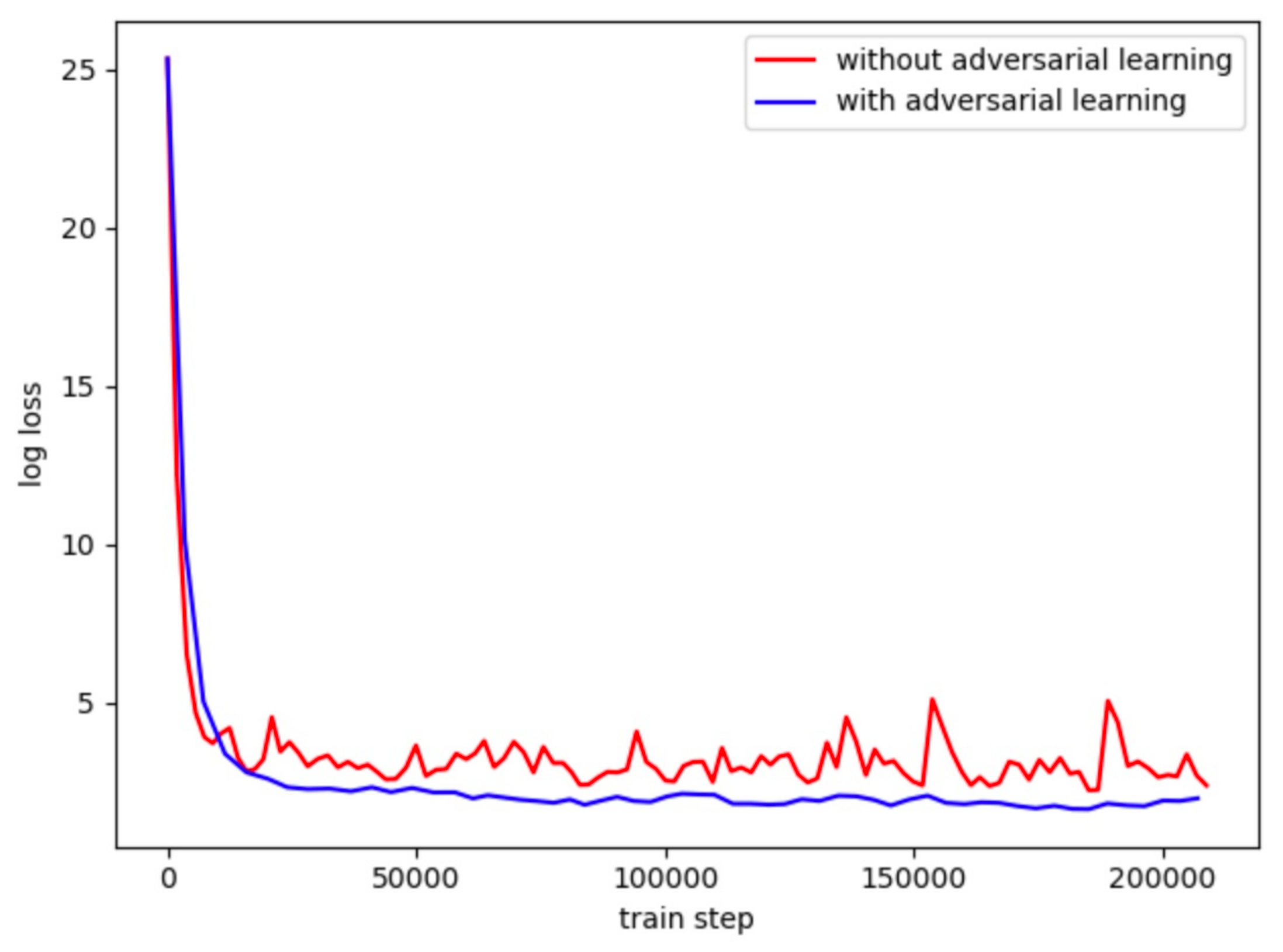}
\caption{The development of the Semantic Matching Loss (i.e. Eq.\ref{eq.logloss}) during the two training procedures (i.e. with adversarial learning $vs.$ without adversarial learning).} 
\label{pic.adv_loss}
\end{figure}

\begin{figure}
\subfigure[]{
    \includegraphics[width=1.55in]{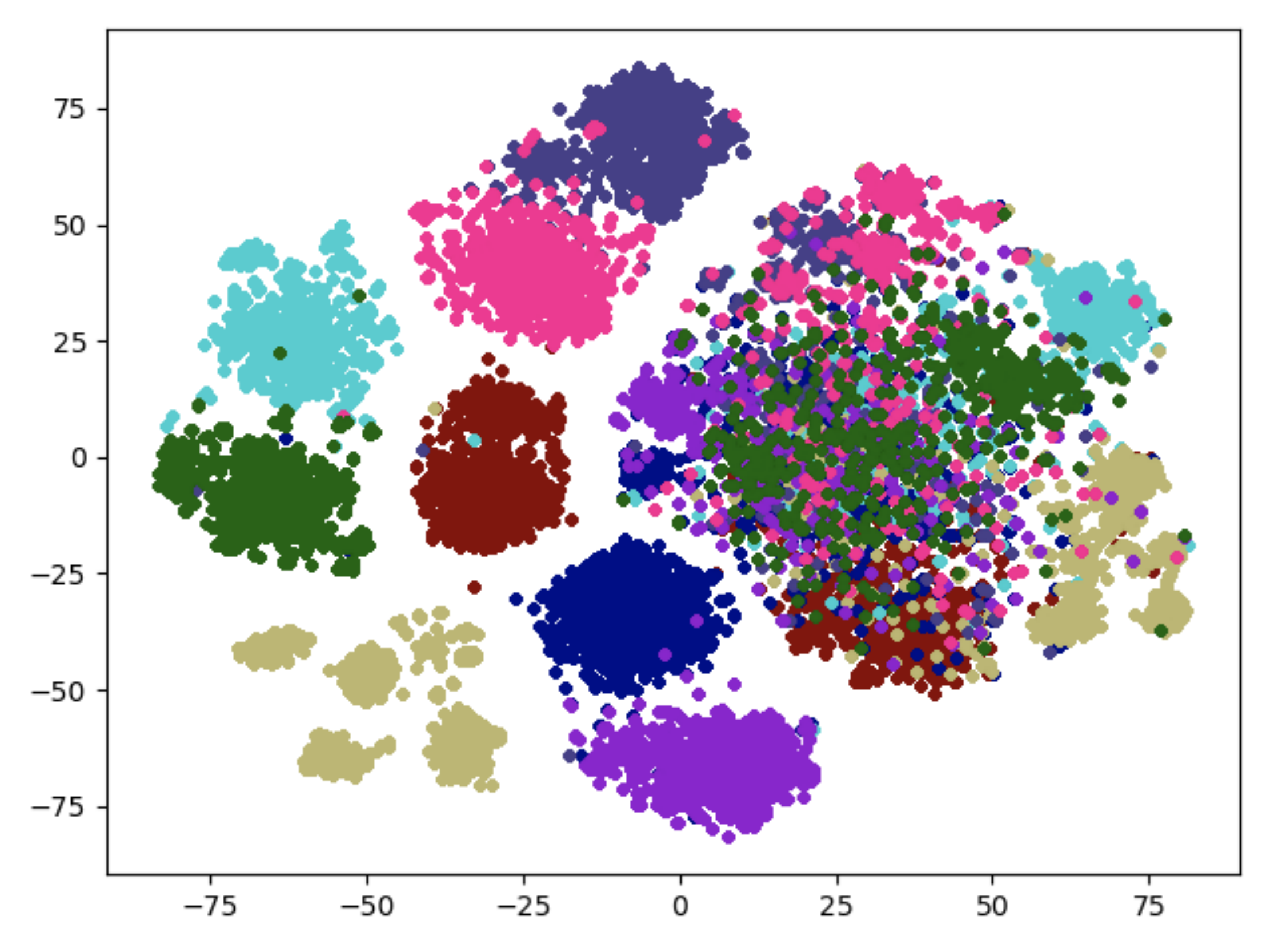}
    \label{tsne_before}
}
\subfigure[]{
    \includegraphics[width=1.55in]{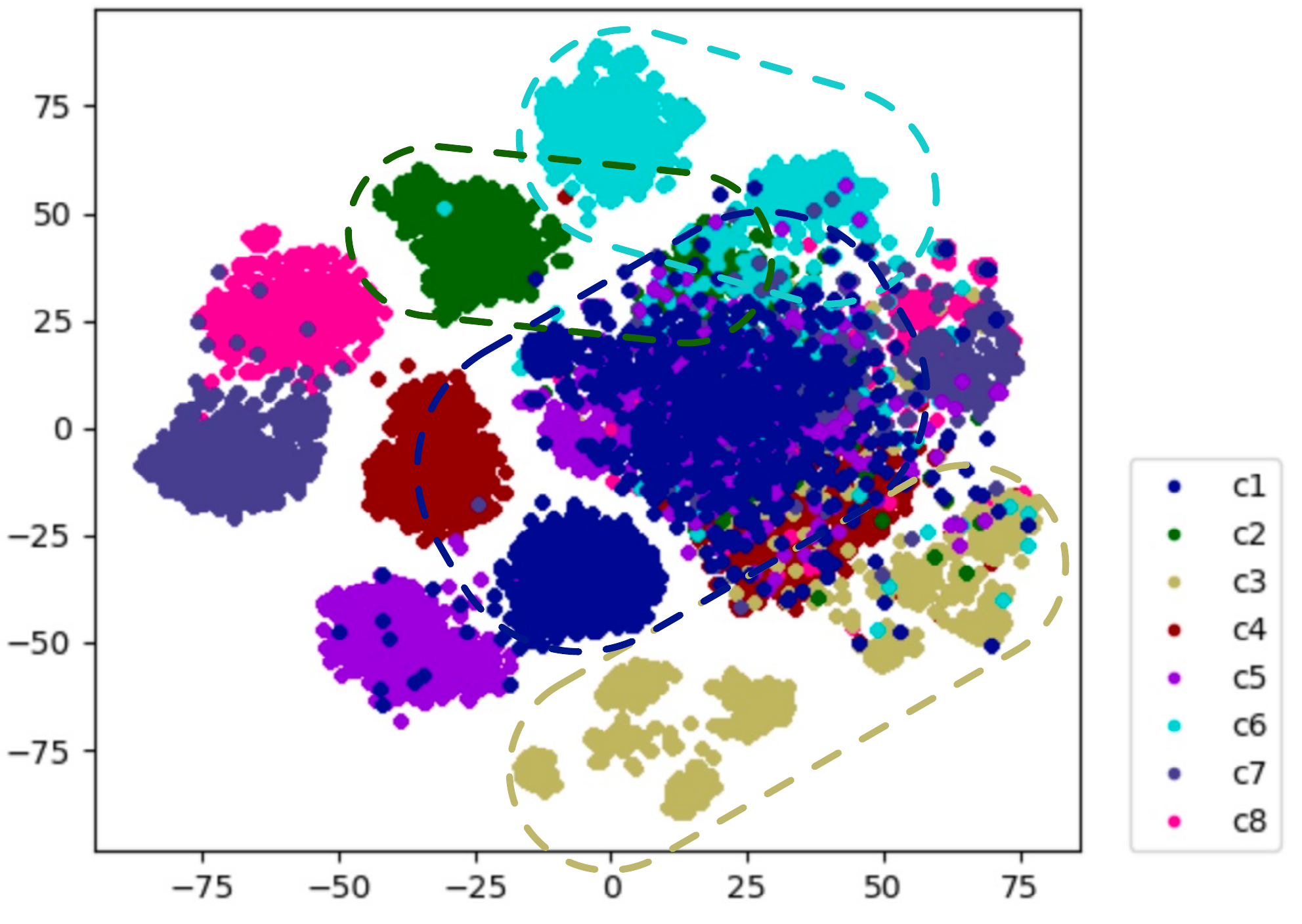}
    \label{tsne_after}
}
\caption{The t-SNE visualization: (a) without adversarial learning, (b) with adversarial learning and the points in the dotted bounding boxes represent they are close each other. The color denotes the different categories.} 
\label{tsne} 
\end{figure}

Fig.\ref{pic.adv_loss} shows the real value of the semantic matching loss (i.e., Eq.\ref{eq.logloss}) during the training procedure. We can find that the values of the loss with the adversarial learning converges smoothly and has a lower loss. To further explore the effectiveness of the adversarial learning, we visualize the distribution of query and product embeddings returned by our trained model through t-SNE\cite{van2008visualizing} tool. In detail, we sample 8000 points for query and product, respectively. These points belong to 8 specific categories. As shown in Fig.\ref{tsne_after}, it is glaringly obvious that the points with the same color (n the dotted bounding boxes) are closer than those points in Fig.\ref{tsne_before}. This result is in accordance with the offline result which indicates the adversarial learning helps to keep the samples from different domains well aligned. 

\subsection{Online System Deployment}
Serving over 100 million cross-border active users is not a trivial task. In deployment, we encounter some challenges. One of the main concern is to balance between complexity of the multi-layer transformers and online computational efficiency. The industry applications always have a strict constraints on the response time of models deployed in production. In order to perform the approximate near neighbor (ANN) with low latency, we pre-compute the embeddings of the hot queries (i.e., account for nearly 87\% search traffic in our E-commerce platform) and all the products. These embeddings are indexed in a distributed database for serving.
At run-time, only the tail queries which can not be cached offline still need to be computed online with high latency. Thus, we propose the compression method to minimize the serving cost. In detail, we compress the layer number of the multi-layer transformers on query side with the knowledge distillation while remain constant on product side. Thanks to it, the computational performance increased by more than 2 times.

\section{Related Work}

\vspace{2mm} \noindent \textbf{Pre-training:} Pre-training is a technology to learn models by leverage information from other related tasks. This technique has been widely adopted in computer vision (CV) at first. With the success of AlexNet \cite{krizhevsky2012imagenet} in ImageNet\cite{deng2009imagenet} classification, most of the researchers subsequently found that the CNN-based neural networks\cite{simonyan2014very,xie2017aggregated} pre-trained on the large-scale image corpus can server well as generic feature representation for a variety of down-stream tasks\cite{donahue2014decaf}. The development of the pre-training in Natural Language Processing (NLP) is behind in CV. In recent years, there are research works\cite{radford2018improving,yang2019xlnet} on pre-training the generic representation. BERT\cite{devlin2018bert}, a Transformer-based representation model for NLP. It learns a universal language encoder by pre-training with large-scale unlabeled data. 

\vspace{2mm} \noindent \textbf{Cross-modal Retrieval:} The cross-modal retrieval methods aim to measure the similarity between the samples from different modalities directly in a common subspace. 
In this paper, we focus on the real-valued representation learning. The methods of Canonical Correlation Analysis (CCA), Deep CCA\cite{andrew2013deep} and variants of CCA-based methods \cite{shao2016deep} are unsupervised methods. These methods maximize the correlations to learn common representations for different types of data. 
In addition, generative adversarial networks (GANs) \cite{goodfellow2014generative} have been proposed in cross-modal retrieval. In these methods\cite{wang2017adversarial,xu2019deep}, a discriminator is introduced to distinguish samples from different modalities. Meanwhile, the studies\cite{chi2018dual,xu2018modal} also apply GAN in zero-shot cross-modal retrieval task. 

Recently, researchers extended the BERT for cross-modal task and achieved great success. The latest BERT-based visual-linguistic models include VL-BERT\cite{su2019vl}, ViLBERT\cite{lu2019vilbert}, VisualBERT\cite{li2019visualbert}, Unicoder -VL\cite{li2020unicoder}. In detail, these algorithms use RoI method which extracts RoIs from images as image tokens. FashionBERT\cite{gao2020fashionbert} adopts patch method, which splits image into patches as image tokens. However, different from the images in the general domains, product images in E-commerce context usually contain only one object. This makes patch method and RoI method inefficient for extracting information from the product image. 

\section{Conclusion}
In this paper, we focus on the query text and product matching in cross-modal retrieval of the E-commerce. There are long term benefits by introducing the multiple modalities into the E-commerce retrieval to boost the performance. We propose ACE-BERT, a novel adversarial cross-modal enhanced BERT for E-commerce retrieval. In particular, we use the RoI-based patch method to extract the patch-level feature and incorporate the pixel-level feature into the BERT backbone. Furthermore, we adopt the adversarial learning to bridge the representation gap across domains and incorporate the additional hot query to further improve the retrieval performance. We systematically carry out the online A/B tests to verify the feasibility of the algorithm.

\bibliographystyle{ACM-Reference-Format}
\bibliography{main}


\end{document}